\documentclass[letterpaper]{JHEP3}

\usepackage{epsfig}

\def\Box{{\hbox{$\sqcup$}\llap{\hbox{$\sqcap$}}}}

\def\be{\begin{equation}}
\def\ee{\end{equation}}
\def\bea{\begin{eqnarray}}
\def\eea{\end{eqnarray}}
\def\nn{\nonumber}

\def\exd{{\rm d}}

\def\endignore{}
\def\ignore #1\endignore{} 

\def\bd{\begin{displaymath}}
\def\ed{\end{diplaymath}}

\def\ie{{\it i.e.}}

\def\cW{{\cal W}}
\def\cA{{\cal A}}

\title{General Axisymmetric Solutions and Self-Tuning in  6D Chiral
  Gauged Supergravity}

\author{  C.P.~Burgess~$^1$, F.~Quevedo~$^2$,  G.~Tasinato,$^3$ and
   I.~Zavala$^4$
\\
$^1$ Physics Department, McGill University, 3600 University Street,
Montr\'eal, Qu\'ebec, Canada, H3A 2T8. \\
$^2$ Centre for Mathematical Sciences, DAMTP, Cambridge
University, Cambridge\\ CB3 0WA UK.\\
  $^3$ Physikalisches Institut der Universit\"at Bonn,
    Nussallee 12, 53115 Bonn, Germany.\\
$^4$ Department of Physics, University of Colorado,
    Boulder, CO, 80309, USA. \\}

\date{}

\abstract{We re-examine the properties of the axially-symmetric
solutions to chiral gauged 6D supergravity, recently found in refs.~{\tt
hep-th/0307238} and {\tt hep-th/0308064}. Ref.~{\tt
hep-th/0307238} finds the most general solutions having two
singularities which are maximally-symmetric in the large 4
dimensions and which are axially-symmetric in the internal
dimensions. We show that not all of these solutions have purely
conical singularities at the brane positions, and that not all
singularities can be interpreted as being the bulk geometry
sourced by neutral 3-branes. The subset of solutions for which the
metric singularities are conical precisely agree with the
solutions of ref.~{\tt hep-th/0308064}. Establishing this
connection between the solutions of these two references resolves
a minor conflict concerning whether or not the tensions of the
resulting branes must be negative. The tensions can be both
negative and positive depending on the choice of parameters. We
discuss the physical interpretation of the non-conical solutions,
including their significance for the proposal for using
6-dimensional self-tuning to understand the small size of the
observed vacuum energy. In passing we briefly comment on a recent
paper by Garriga and Porrati which criticizes the realization of
self-tuning in 6D supergravity.}

\preprint{McGill-04/17, DAMTP-2004-75, \\ COLO-HEP-501}


\begin{document}

\section{Introduction}

Chiral gauged six-dimensional supergravity
\cite{6dsugra1,6dsugra2} has seen considerable study for more than
two decades. This has happened partly due to its sharing many of
the features of ten-dimensional supergravity --- and so also of
string vacua --- such as the existence of chiral fermions
\cite{6dsugra1,chiralfermions} with nontrivial Green-Schwarz
anomaly cancellation \cite{6dsugra1,6dac} as well as the
possibility of having chiral compactifications down to flat 4
dimensions \cite{ss}. Part of the attention has also been due to
the relative simplicity of the 6D theories, which makes them
comparatively simple places to explore some of the ideas
\cite{abpq1} (such as compactifications with fluxes \cite{fluxes})
which have arisen in studies of 4D vacua of their
higher-dimensional cousins.

More recently, interest has also come from the proposal that 6D
supergravity might provide a viable example of the self-tuning of
the effective 4D cosmological constant
\cite{Towards,Update}.\footnote{See  ref.~\cite{GP} for a
dissenting point of view, on which we comment in the Appendix.}
The proposed self-tuning mechanism --- the validity of which is
still under active study --- comes in two parts. The first part of
the mechanism involves the cancellation of arbitrary brane
tensions by the classical response of the bulk supergravity
degrees of freedom, while the second part involves the size of the
quantum corrections to the classical response due to bulk loops.
Our discussion here is relevant for the first part of the
argument. We have nothing to say in the present paper concerning
the quantum part of the proposal.

Although the original claim for the classical part of this
argument was based on earlier arguments from non-supersymmetric 6D
theories \cite{CLP}, more direct conclusions have in some cases
since become possible due to the subsequent discovery of a broad
class of solutions to the 6D supergravity equations
\cite{GGP,Warped}. In particular, ref.~\cite{GGP} find the most
general solution subject to the assumptions of $(i)$ maximal
symmetry in the large 4 dimensions; and $(ii)$ axial symmetry in
the compact 2 dimensions. For all of the solutions within this
category the intrinsic geometry of the large 4 dimensions is found
to be flat.

In this note we examine further the properties of these solutions,
with the immediate goal of more directly establishing the features
of the metric singularities which most of these solutions have. In
particular, if we interpret these singularities as being due to
the gravitational back-reaction of 3-branes we wish to ascertain
how the properties of the bulk field configurations are related to
the properties (such as tensions) of these branes. In particular,
we discuss the relevance of our results to the validity of the
classical part of the self-tuning argument (in a sense we outline
more concretely in what follows). In passing we resolve a minor
difference between ref.~\cite{Warped}, which allows single-brane
solutions having positive tension \footnote{Configurations involving
only positive tension branes
  have also been found recently
in 6D gravity coupled to a sigma model \cite{seif}.}, and v3 of ref.~\cite{GGP},
which states all of these solutions must involve negative tension.

Our study leads to the following general conclusions:
\begin{enumerate}
\item The 3-parameter set of general solutions given in
ref.~\cite{GGP} contain examples for which the metric
singularities are {\it not} purely conical, and these come in two
categories. Some can be interpreted as describing the fields of
two localized 3-brane like objects, while others are better
interpreted as the bulk fields which are sourced by a combination
of a 3-brane and 4-brane, rather than being due to two 3-branes;
\item The 2-parameter subset of the general solutions of
ref.~\cite{GGP} which have purely conical singularities can be
interpreted as being sourced by two 3-branes, and these solutions
precisely agree with the solutions found in ref.~\cite{Warped}. In
particular, the tensions, $T_\pm$, of the two 3-branes which
source all such solutions must satisfy the relation $f(T_+,T_-) =
0$, derived in \cite{Warped}, and which is given explicitly by
eq.~(\ref{Tconstraint});
\item Although topological constraints do restrict the kinds of
tensions which are possible (for those solutions having only
conical singularities), they need not prevent choosing both
tensions to be positive if desired. This observation can be
important for the analysis of the stability of these geometries.
\item We find that it is locally possible to determine the
parameters of the general solution in terms of the two source
brane tensions, even in the case where the singularities are not
conical. This shows that a solution very likely exists for any
pair of tensions, at least within a neighborhood of the conical
solutions (and possibly for generic tensions). This has
implications for the question of how the bulk adjusts if the
tensions of the source branes are adjusted independently, as might
occur for instance if a phase transition were to take place on one
of the branes.
\end{enumerate}

In the next section, \S2, we describe the general solutions of
ref.~\cite{GGP}, and outline some of their properties. In
particular we here identify the nature of their singularities. \S3
is dedicated to analyzing the properties of the configurations
with purely conical singularities. \S4 then discusses the physical
interpretation of those solutions whose singularities are more
complicated than conical, and finds the tension of the source
branes as functions of the parameters of the solution in the
general case. In \S5 we show that the subset of solutions having
only conical singularities precisely agrees with those found in
ref.~\cite{Warped}, and \S6 provides a discussion of the relevance
of our results for the 6D self-tuning proposal. We close, in an
appendix, with a brief critique of the arguments of Garriga and
Porrati \cite{GP}, who criticize the possibility of realizing
self-tuning in 6 dimensions.

\section{The Solutions of Ref.~\cite{GGP}}

In this section we describe the general solution given in the
appendix of ref.~\cite{GGP}. Our goal in doing so is to identify
the nature of the singularities of the space-times, and to
establish a baseline for the later comparison with the solutions
of ref.~\cite{Warped}.

The field content of 6D supergravity which is nonzero for these
solutions consist of the 6D metric, $g_{MN}$, a 6D scalar dilaton,
$\phi$ and a $U(1)$ gauge potential, $A_M$. The field equations
which these fields must solve are\footnote{For ease of comparison
we here adopt the conventions of ref.~\cite{GGP}, which differ
from those of refs.~\cite{Towards} and \cite{Warped} in ways which
are spelled out in detail in subsequent sections.}
\bea \label{E:Beom}
    &&\Box \, \phi  - \frac14 \, e^{\phi/2} \; F_{MN} F^{MN} + 8 g^2
    e^{-\phi/2} = 0 \nn\\
    &&D_M \Bigl( e^{\phi/2} \, F^{MN} \Bigr) = 0  \\
    &&R_{MN} - \frac14 \, \partial_M\phi \, \partial_N \phi -
    \frac12 \,
    e^{\phi/2} \, F_{MP} {F_N}^P - \frac14 \, (\Box \phi )\, g_{MN} =
    0 , \nn
\eea
where $g$ is the gauge coupling constant for a specific abelian
$R$-symmetry, $U(1)_R$, within the 6D theory.

For later purposes, we remark that these equations are invariant
under the constant classical scaling symmetry,
\be \label{rescalings}
    g_{MN} \to \xi \, g_{MN} \qquad \hbox{and} \qquad
    e^{\phi/2} \to \xi \, e^{\phi/2} \,.
\ee
Furthermore, although the scalings $e^{\phi/2} \to \zeta^2 \,
e^{\phi/2}$ and $F_{MN}\to \zeta^{-1} \, F_{MN}$ are not a
symmetry, they have the sole effect of scaling $g \to g' =
\zeta\, g$.

\subsection{Solutions for General $\lambda_3$}

Ref.~\cite{GGP} construct the most general solution to these
equations for which 4 Lorentzian dimensions are maximally
symmetric, and for which the internal 2 dimensions are axially
symmetric. The metric for these solutions has at most two
singularities in the extra dimensions. Their general solution is
given explicitly by
\bea
    e^\phi &=& W^{4} \, e^{2 \lambda_3 \eta} \nn \\
    F &=& \exd A =
\left( \frac{q\,a^2}{ W^{2}}\, \right) e^{- \lambda_3 \eta}\,
    \exd \eta\wedge \exd \psi
    \label{ggp300}\\
    \exd s^{2} &=& W^{2} \eta_{\mu \nu} \, \exd x^{\mu} \exd x^{\nu}
    +a^{2} W^8 \exd \eta^{2} +a^{2} \exd \psi^{2} \,,\nn
\eea
where
\bea
    W^{4} &=& \left( \frac{q \lambda_2}{4g\lambda_1} \right)
    \frac{\cosh\left[\lambda_1(\eta-\eta_1)\right]}{\cosh
    \left[\lambda_2(\eta-\eta_2)\right]}\\
    a^{-4} &=& \left( \frac{g q^{3}}{\lambda_1^{3}\lambda_2}
    \right) e^{-2\lambda_3
    \eta} \cosh^{3}\left[\lambda_1(\eta-\eta_1)\right] \cosh
    \left[\lambda_2(\eta-\eta_2)\right] \,.
\eea
Here $q$, $\eta_a, a=1,2$ and $\lambda_i, i=1,2,3$ are constants,
and the parameters $\lambda_i$ satisfy the constraint
\be\label{constrlamb}
    \lambda_1^{2}+\lambda_3^{2}=\lambda_2^2\,.
\ee
Following \cite{GGP} we take $\lambda_1$ and $\lambda_2$ to be
positive, and so the last relation implies the inequality
$\lambda_2 \ge \lambda_1$.

At face value we have a total of 5 integration constants:  $q$,
$\lambda_1$, $\lambda_2$, $\eta_1$ and $\eta_2$. However one
combination of these five constants corresponds to the scaling
symmetry, eq.~(\ref{rescalings}), whose action on the solution may
be represented by the following transformations:
\be
    \psi \to \xi^{2} \psi, \quad q \to \xi^{2} q  \,.
\ee
A second combination similarly corresponds to the second rescaling
discussed above, whose sole effect is to change $g$ to $\zeta\, g$,
and which for $\lambda_3 \ne 0$ is represented on the solutions by
\be
    q \to \zeta \, q, \quad  \eta \to \eta +
    \frac{2}{\lambda_3} \, \ln \zeta , \quad \eta_i \to \eta_i
     + \frac{2}{\lambda_3} \ln \zeta  \,,
\ee
leaving a total of 3 nontrivial parameters. (If $\lambda_3 = 0$
then one integration constant -- say $\eta_1$ -- can similarly be
removed by shifting the coordinate $\eta$ without also rescaling
$q$.)

\subsection{Singularities}

Singularities of the metric can occur where its components vanish
or diverge. Inspection of eq.~(\ref{ggp300}) shows that this only
occurs when $\eta \to -\infty$ and $\eta \to +\infty$. In these
limits $W \propto \exp \left[ \frac14 \, (\lambda_1 -
\lambda_2)|\eta| \right]$, and so clearly $W$ remains bounded
because $\lambda_1 \le \lambda_2$, and vanishes if $\lambda_1$ is
strictly smaller than $\lambda_2$. Also, $a \propto \exp \left[
\frac14 \, (2 \lambda_3 \eta - (3 \lambda_1 + \lambda_2)|\eta| )
\right]$ in the limit of large $|\eta|$, and so if $\pm 2\lambda_3
+ (3\lambda_1 + \lambda_2)$ is nonzero then $a$ either diverges or
vanishes as $\eta \to \pm \infty$.

To examine the nature of the metric singularities in more detail
at these points we consider the internal two-dimensional metric,
which has the form $\exd s^2_2 = a^2 W^8 \exd \eta^2 + a^2 \exd
\psi^2$. In general, in the vicinity of a conical singularity a 2D
metric can always be re-written in the form
\be\label{consin}
    \exd s_2^{2}= \exd r^{2}+a_0^{2} r^{2} \exd \psi^{2}
\ee
where $\psi$ has period $2\pi$, and the conical deficit angle at
$r = 0$ is given in terms of the constant $a_0$ by $\delta = 2 \pi
(1 - a_0)$. We now show that this form cannot be obtained for
those solutions of ref.~\cite{GGP} for which the parameter
$\lambda_3$ is nonzero. The metric of the solution may be compared
with this form after performing a change of variable from $\eta$
to $r$, using $a W^4 \exd \eta = \exd r$.

In the limit $\eta \to \pm\infty$ we have
\be \label{chanofvar}
    \exd r =  \frac12 \, \left( \frac{q\,e^{\mp \lambda_1 \eta_1}}{\lambda_1}
    \right)^{1/4}
    \left( \frac{\lambda_2\,e^{\pm \lambda_2 \eta_2}}{g} \right)^{5/4}
    \exp \left[ -\frac14 \,
    (\pm 2 \lambda_3 + 5 \lambda_2 - \lambda_1)  \,
    |\eta| \right] \exd \eta
    \,,
\ee
and so the behavior of $r$ is determined by the sign of
$\Delta_\pm \equiv \pm 2 \lambda_3 + 5 \lambda_2 - \lambda_1$,
which must be positive if $\lambda_2^2 = \lambda_1^2 +
\lambda_3^2$.\footnote{To see this one can proceed in the
following way. We have to show that  $5 \lambda_2 \ge \lambda_1
\mp 2 \lambda_3$. If the right hand side of the inequality  is
negative, the inequality is satisfied since $\lambda_2$ is
positive. If the right hand side is positive, we can square the
inequality, obtaining $25(\lambda_1^2 + \lambda_3^2) \ge
(\lambda_1 \mp 2 \lambda_3)^2$. Equivalently, this implies $(2
\lambda_1 \pm \lambda_3)^2 + 20(\lambda_1^2 + \lambda_3^2) \ge 0$,
which is always satisfied for real $\lambda_i$.} It follows that
$r \to 0$ as $\eta \to \pm\infty$. As $\eta \to \pm \infty$ the
metric functions $W$ and $a$ behave as $W \propto r^{\omega_\pm}$
and $a \propto r^{\alpha_\pm}$, with the powers $\alpha_\pm$ and
$\omega_\pm$ satisfying the identity $\alpha_\pm + 4 \omega_\pm =
1$, and being given explicitly by
\be \label{omegaalpha}
    \omega_\pm = \frac{\lambda_2 - \lambda_1}{\Delta_\pm} \,, \qquad \qquad
    \alpha_\pm = \frac{\pm 2 \lambda_3 + \lambda_2 + 3 \lambda_1}{\Delta_\pm} \,.
\ee
Notice that in the special case $\lambda_1 = \lambda_2 \equiv
\lambda$, we have $\lambda_3 = 0$ and so $\omega_\pm = 0$ and
$\alpha_\pm = 1$.

We now turn to the asymptotic behavior of the functions $W$ and
$a$ as $r \to 0$. As mentioned previously, the inequality
$\lambda_2 \ge \lambda_1$ implies $\omega_\pm \ge 0$ and so
(unless $\lambda_1 = \lambda_2$) $W$ vanishes at both
singularities. The behavior of $a$ near the singularities is
similarly controlled by the power $\alpha_\pm$. For instance,
$\alpha_+ = 0$ can happen for positive $\lambda_1$ and $\lambda_2$
only when $\lambda_3 = 2 \lambda_1[ -1 - 1/\sqrt3] < 0$ and
$\lambda_2 = \lambda_1 [1 + 4/\sqrt3] > 0$. Similarly, $\alpha_- =
0$ requires $\lambda_3 = 2 \lambda_1 [1 + 1/\sqrt3] > 0$ and
$\lambda_2 = \lambda_1 [ 1 + 4/\sqrt3] > 0$. Thus, the
inequalities $\alpha_+ < 0$ and $\alpha_- < 0$ describe the two
wedges in the upper-half $\lambda_1 - \lambda_3$ plane illustrated
in Fig.~(\ref{sbrpen}). It is clear from this that it is possible
to choose non-negative $\lambda_1$ and $\lambda_2$ to ensure one
of two possibilities: ($i$) $\alpha_\pm$ both positive; or ($ii$)
$\alpha_\pm$ have opposite sign. It is not possible for both
$\alpha_\pm$ to be negative, and both can vanish only if
$\lambda_1 = \lambda_2 = 0$.

\FIGURE[ht]{
\let\picnaturalsize=N
\def\picsize{4.0in}
\def\picfilename{regions.eps}
\ifx\nopictures Y\else{\ifx\epsfloaded Y\else\input epsf \fi
\let\epsfloaded=Y
\centerline{\ifx\picnaturalsize N\epsfxsize \picsize\fi \epsfbox{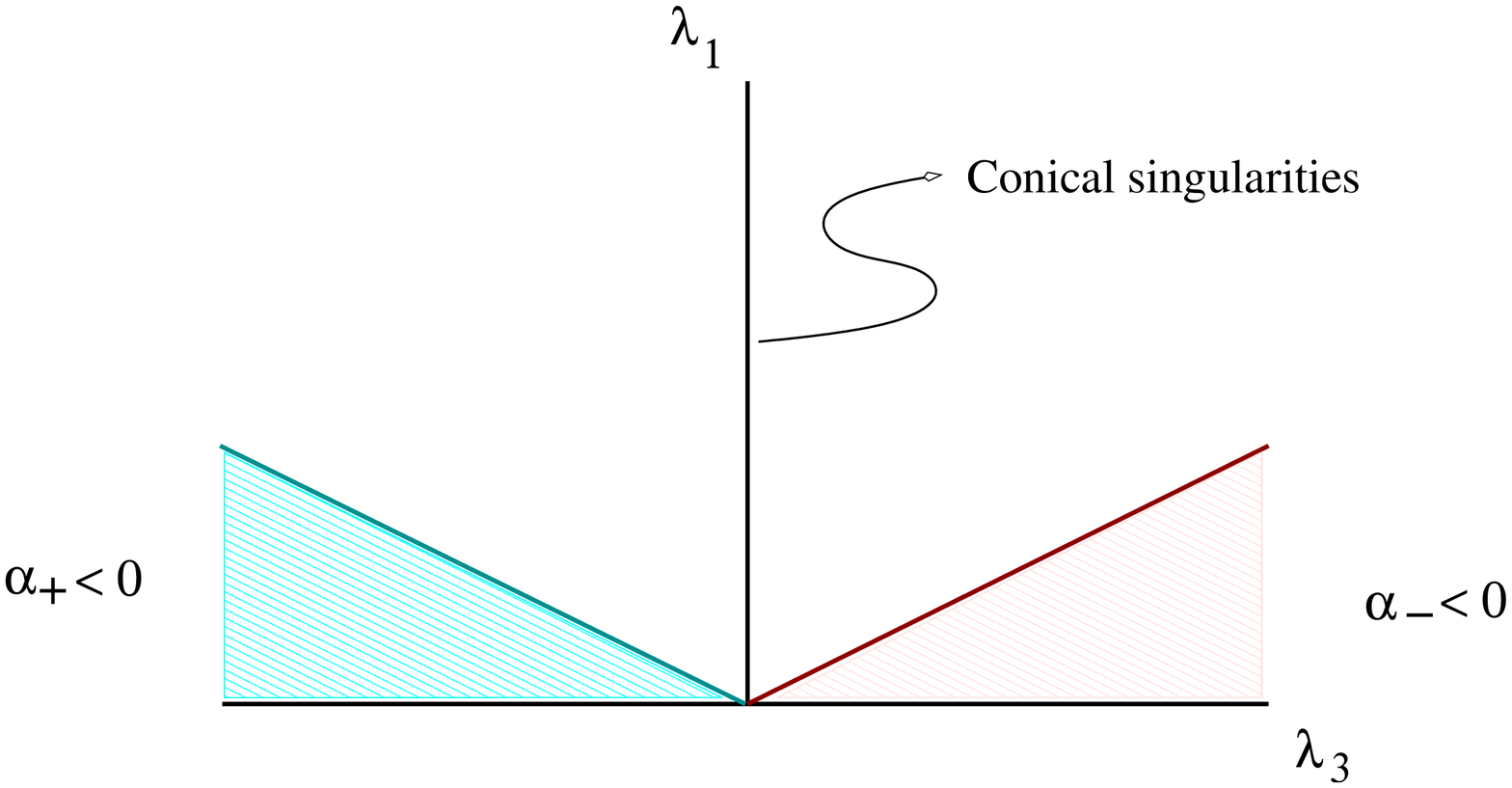}}}\fi
\caption{Solid regions indicate where in the $\lambda_1 -
\lambda_3$ plane,  the powers $\alpha_\pm$ are negative.
Notice the existence of a big  region for which  both
$\alpha_\pm > 0$. Only for $\lambda_3=0$ are the singularities
conical.
\label{sbrpen}}}

With these results we see that the angular part of the metric near
the singularities becomes $a^2 \exd \psi^2 \to a^2_\pm \,
r^{2\alpha_\pm} \, \exd \psi^2$, where $a_\pm$ is a constant. In
the special case $\lambda_3 = 0$ (and so also $\lambda_1 =
\lambda_2 \equiv \lambda$) we have $\alpha_\pm = 1$ and
$\omega_\pm = 0$, and so the metric takes a conical form with
defect parameter $a_\pm = \lambda (4g/q) e^{\pm\lambda ( \eta_1 -
\eta_2)}$ (corresponding to the defect angle $\delta_\pm = 2 \pi
(1- a_\pm )$). For any other $\alpha_\pm$ the 2D metric {\it
cannot} be brought into the conical-defect form near the
singularity, and so the metric here is not locally a cone.

Notice also that so long as $\alpha \ne 1$ local bulk-curvature
invariants diverge in the limit $r \to 0$ (by contrast with the
delta-function behavior which obtains in the case of a conical
singularity). A similar statement also holds for the bulk dilaton
and electromagnetic fields, which can diverge or vanish at the
positions of the branes depending on the particular values chosen
for $\lambda_1$ and $\lambda_2$. We further discuss the
interpretation of the singularities when $\alpha_\pm \ne 1$ in
\S3, below.

\section{The Special Case $\lambda_3 = 0$}  \label{sec:properties}

In this section, we analyze the properties of general GGP
solutions for the special case $\lambda_3 = 0$, for which the
above discussion shows that the metric singularities are conical.
Because the parameters $\lambda_1$ and $\lambda_2$ are positive
and satisfy the relation (\ref{constrlamb}), we must in this case
also choose $\lambda_1 = \lambda_2 = \lambda$. We shall see that
in the particular case $\lambda = 1$, one of the conical
singularities disappears, leaving only a single singularity. This
is the case considered in Section 2 of
ref.~\cite{GGP}.\footnote{Indeed, in the case of one singularity
ref.~\cite{GGP} proves this solution is the most general possible
{\it without} making the assumption of axial symmetry in the extra
dimensions.}

\subsection{The Solution}

When $\lambda_3 = 0$ and $\lambda_1 = \lambda_2 = \lambda$ the
general solution reduces to


%
\bea
    e^\phi &=& W^{4} \nn \\
    F &=& \left( \frac{q\,a^2}{ W^{2}} \right)
    \, \exd \eta\wedge \exd \psi
    \label{ggp301}\\
    \exd s^{2} &=& e^{{\phi}/{2}} \eta_{\mu \nu} \,
    \exd x^{\mu} \exd x^{\nu}
    + a^{2} W^8 \exd \eta^{2} +a^{2} \exd \psi^{2} \,,\nn
\eea
where
\bea  \label{genWformula}
    W^{4} &=& \left( \frac{q}{4g} \right)
    \frac{\cosh\left[\lambda(\eta-\eta_1)\right]}{\cosh
    \left[\lambda(\eta-\eta_2)\right]}\\
    a^{-4} &=& \left( \frac{g q^{3}}{\lambda^{4}} \right)
    \cosh^{3}\left[\lambda(\eta-\eta_1)\right]
    \cosh
    \left[\lambda(\eta-\eta_2)\right] \,.
\eea
We are left with the 4 integration constants, $q$, $\lambda$,
$\eta_1$ and $\eta_2$, and one of $\eta_1$ or $\eta_2$ can be
removed by appropriately shifting $\eta$. As previously discussed
the scale invariance, (\ref{rescalings}), is implemented by $q \to
\xi^2 \, q$ and $\psi \to \xi^2 \, \psi$. The authors of
ref.~\cite{GGP} remove this classical symmetry by using it to
impose the condition
\be \label{etadiff}
    e^{\lambda(\eta_1-\eta_2)} = \frac{r_1}{r_0}\,\,,
\ee
where the $r_i$ are given by
\be
    r_0^2=\frac{1}{2g^2}\,\hskip0.5cm,\hskip0.5cm r_1^{2}=\frac{8}{q^2} \,.
    \label{exprpar}
\ee
This choice, and the freedom to shift $\eta$, reduces the number
of integration constants to two.

At this point, it is useful to define a new radial coordinate $0
\le \sigma < \infty$, by:
\be
    \sigma=r_1 e^{\lambda(\eta-\eta_1)} \,,
\ee
in which case the solution, eq.~(\ref{ggp301}), becomes
\bea
    \phi &=& \ln \left( {\frac{f_1}{f_0}} \right) \nn \\
    F &=& \left( \frac{q \lambda \,\sigma}{f_{1}^{2}} \right)
    \, \exd\sigma \wedge \exd\psi \label{ggp302}\\
    \exd s_{6}^{2} &=& \left(\frac{f_1}{f_0}\right)^{{1}/{2}}
    \left( \eta_{\mu \nu} \, \exd x^{\mu} \exd x^{\nu} \right)+
    \left(\frac{f_1}{f_0}\right)^{{1}/{2}} \left(
    \frac{\exd \sigma^{2}}{f_0^{2}}+
    \frac{\lambda^{2} \sigma^{2}}{f_1^{2}} \, \exd \psi^{2}
    \right) \,, \nn
\eea
where
\be
    f_0 \equiv 1+\frac{\sigma^{2}}{r_0^{2}} \hskip 1 cm \hbox{and}
    \hskip 1 cm f_1
    \equiv 1+\frac{\sigma^{2}}{r_1^{2}}\,.
\ee
We see that in this form the solution explicitly depends only on
the two independent integration constants, which we may take to be
$q$ and $\lambda$.

Analyzing the metric of eq.~(\ref{ggp302}), it is easy to see that
for $\lambda \ne 1$ there are two conical singularities, one at
$\sigma=0$ and the other at $\sigma=+\infty$. The corresponding
deficit angles are given by
\be\label{defdefan}
    \frac{\delta_0}{2\pi} =1-\lambda \hskip 2cm
    \frac{\delta_{\infty}}{2\pi}
    =1-\lambda\left(
    \frac{r_1^{2}}{r_0^{2}} \right) \,.
\ee
As is clear from these expressions, the conical singularity at the
origin disappears when $\lambda=1$. The tensions of the branes at
the conical singularities may be read off from these expressions,
since they are proportional to the deficit angles:
\be \label{Tvsdelta}
     T = \frac{\delta}{8\pi G} \,,
\ee
where we briefly reinstate the 6D Newton constant, $8 \pi G =
\kappa^2$.

\subsection{Topological Constraints}\label{tc1}

In this subsection, we re-examine the topological constraints that
the previous solution must satisfy. Our discussion here, which
generalizes the topological constraints discussed for the simplest
unwarped solution in refs.~\cite{Towards,GibbonsPope} (see also
\cite{ignacio}), follows the method presented in \cite{GGP}. Our
interest is mainly to investigate the consequences of this
constraint for the sign of the brane tensions.

To this end we notice that the field strength $F$ of the solution,
eq.~(\ref{ggp302}), locally can be written in terms of the
one-form potential
\be\label{AAA}
    A = - \left( \frac{4 \lambda}{q f_1} \right) \exd \psi\,.
\ee
In order for this section of the gauge bundle to be globally
well-behaved it is necessary to impose the quantization condition
\be\label{quaco}
    \frac{4 \lambda\, \tilde{g}}{q}  = N\,,
\ee
where $N$ an integer and $\tilde{g}$ is the 6D gauge coupling for
the particular gauge field which is nonzero in the classical
solution. Notice that by writing only one quantization condition
we assume the background $U(1)$ to lie completely within a simple
factor of the gauge group for which there is only a single gauge
coupling.\footnote{The more general case where the background
field overlaps two gauge generators having different couplings is
considered in ref.~\cite{GibbonsPope}.} Even so, in general
$\tilde{g}$ need not equal the coupling, $g$, appearing in the
dilaton potential since the field strength which is turned on in
the classical solution need not be for the specific $U(1)_R$
generator whose coupling is $g$.

\medskip\noindent {\bf The Special Case $g = \tilde{g}$.}\medskip

\noindent If we do take $g = \tilde{g}$ --- such as we must if the
background gauge field gauges the $U(1)_R$ symmetry of 6D chiral
supergravity --- it is immediate that at least one of the brane
tensions must be negative so long as $N \ne 0$ \cite{GGP}. To see
this we use the quantization result, (\ref{quaco}), in the form
\be
    \frac{r_1}{r_0} = \frac{4 g}{q} = \frac{N}{\lambda}
    \label{quantfc} \,,
\ee
to see that the deficit angles become
\be
    \frac{\delta_0}{2\pi} =1-\lambda \hskip 1cm \hbox{and} \hskip
    1cm
    \frac{\delta_{\infty}}{2\pi} =1-\frac{N^{2}}{\lambda} \,.
\ee
{}From these expressions it is easy to see that at least one of
the branes must have negative tension, since $\lambda$ is positive
and $N^2 \ge 1$.

An interesting particular case of this situation is obtained by
specializing to the {\it rugby-ball} geometry, for which the two
tensions are equal \cite{Towards,GibbonsPope}. Using
eq.~(\ref{defdefan}) we see this corresponds to the special case
$r_1=r_0$, for which the following change of radial variable
\be
    \sigma= r_0 \tan{\frac{\theta}{2}}\,,
\ee
allows the metric for the two internal dimensions, taken from the
solution eq.~(\ref{ggp302}), to be rewritten as
\be\label{defect}
    \exd s_2^{2}\equiv
    \left(\frac{f_1}{f_0}\right)^{{1}/{2}} \left( \frac{\exd
    \sigma^{2}}{f_0^{2}}+\frac{\lambda^{2} \sigma^{2}}{f_1^{2}}
    \, \exd
    \psi^{2}
    \right) = \frac{r_0^{2}}{4}  \left(\exd \theta^{2} +\lambda^{2}
    \sin^{2}{\theta} \, \exd\psi^{2} \right) \,.
\ee
{}From this form of the metric it is immediate to see that the
geometry being described is a sphere with deficit angle $\delta =
2\pi(1-\lambda)$ removed at each of the poles. For this metric the
quantization condition, eq.~(\ref{quantfc}), implies $\lambda =
N$, and so the deficit angle $\delta$ of the two conical
singularities is given by
\be
    \frac{\delta}{2\pi}=1-N \,.
\ee
Again the tensions are negative for any nonzero integer $N$.

\medskip \noindent {\bf The Case $g \neq \tilde{g}$.}\medskip

\noindent Situations having positive tension may be found by
taking $g < \tilde{g}$, in which case the quantization condition,
(\ref{quaco}), implies
\be
    \frac{r_1}{r_0} = \frac{Ng}{\lambda \tilde{g}} \,
     \,.
\ee
This leads to the following expressions for the deficit angles:
\be \label{deficitpredictions}
    \frac{\delta_0}{2\pi} =1-\lambda \hskip 2cm
    \frac{\delta_{\infty}}{2\pi}
    =1-\frac{N^{2}}{\lambda} \left( \frac{g}{\tilde{g}} \right)^{2}\,.
\ee
Clearly, both the brane tensions can be positive in this case
provided $\lambda < 1$ and $(g /\tilde{g})^2 < \lambda/N^2$. In
the particular case of the rugby-ball geometry, one finds that the
deficit angle becomes
\be
    \frac{\delta}{2\pi}=1-
    \frac{Ng}{\tilde{g}}\,,
\ee
which can again be positive if $g/\tilde{g} < 1/N$.

Superficially, the above expressions seem to disagree with the
conclusion drawn in version 3 of ref.~\cite{GGP}, wherein it is
claimed that one of the tensions must be negative for any of the
solutions having only one singularity. This contradiction is only
superficial because GGP draw this conclusion under the assumption
that the gauge field turned on in the background is a {\it
nontrivial mixture} of two gauge directions for which the
couplings are different. They do so by taking it to be a linear
combination of an internal gauge group (such as one which commutes
with supersymmetry) and of the specific $U(1)_R$ gauge group whose
gauge coupling, $g$, appears in the dilaton potential. As is clear
from their expressions, the requirement for negative tension
disappears if the background field is purely orthogonal to the
$U(1)_R$ direction, since in this case only one of their two
quantization conditions applies and is in itself insufficient to
force one of the tensions to be negative.

For later convenience we pause to record a relation between the
tensions which always holds when $\lambda_3 = 0$, and which is
obtained from eqs.~(\ref{deficitpredictions}) by eliminating the
common variable $\lambda$ and using the relation (\ref{Tvsdelta}),
$\delta = 8 \pi G \, T$:
\be \label{GGPTensionconstraint}
    (1 - 4 G \, T_0) (1 - 4 G \, T_\infty) = N^2 \left(
    \frac{g^2}{\tilde{g}^2} \right)
    \,.
\ee
This expression generalizes the relation derived in
ref.~\cite{GGP} to the case $\tilde{g} \ne g$.

\section{The Branes Behind the Solutions}

The existence of solutions having non-conical metric singularities
(when $\lambda_3 \ne 0$) raises several interpretational issues,
if these singularities are to be interpreted as the positions of
various branes whose stress-energy and charges source these bulk
fields. In order to better understand what is going on, it is
useful to consider the behavior of the fields in the vicinity of a
single brane for pure gravity (without the dilaton and
electromagnetic fields).

\subsection{Singular Single-Brane Configurations}

Consider therefore pure gravity in 6 dimensions (which also
applies to 6D supergravity in the limit $g = 0$), with the metric
chosen to have the general axially-symmetric form\footnote{The
function $A(r)$ should not be confused with the background value
of the gauge field, as in eq.~ (\ref{AAA}).}
\be \label{bulkgeometry}
    \exd s^2 = W^2(r) \, \eta_{\mu\nu} \, \exd x^\mu \exd x^\nu +
    \exd r^2 + A^2(r) \, \exd\theta^2 \,.
\ee
A simple calculation shows that the condition that this metric be
a vacuum spacetime --- {\it i.e.} one satisfying $R_{MN} = \frac12
\, R \, g_{MN}$ --- has two types of solutions.
\begin{enumerate}
\item The first class of solution is the usual unwarped cone, for
which $W(r) = W_0$ and $A(r) = A_0 \, r$, with $W_0$ and $A_0$
being arbitrary positive constants.
\item The second class of solution has $W(r) = W_0 r^{2/5}$ and
$A(r) = A_0 r^{-3/5}$, where $A_0$ and $W_0$ are again the
integration constants. Unlike the conical metrics, these warped
solutions are not locally flat geometries.
\end{enumerate}
We see in this way the existence of non-conical warped solutions
which are not asymptotically locally-flat, just as for the more
general $\lambda_3 \ne 0$ GGP  solutions. In both Einstein gravity
and the GGP solutions the local curvature invariants diverge as
one approaches the brane position, as opposed to the simple
delta-function behavior which arises for the conical
singularities. Other bulk fields, like the dilaton and Maxwell
fields, can also blow up or vanish at these points depending on
the values chosen for $\lambda_1$ and $\lambda_2$.

The existence of this second type of solution to the Einstein
equations has also been noticed for 4 dimensions in the context of
the gravitational field produced by cosmic strings. However it was
initially discarded on the grounds that it was inappropriate to
the desired cosmic-string applications \cite{Vilenkin}. More
recently it has been re-examined for its possible relevance to
super-massive or global-string configurations \cite{shellard}.
Similar single-brane solutions (involving different powers of $r$
in $W$ and $a$) also exist for the coupled dilaton-Einstein
system, for instance with the result in 4 dimensions being given
in ref.~\cite{gregory}.

\subsection{Interpretational Issues}

In any case, which of these solutions is appropriate in a given
physical situation is determined in principle by a set of boundary
conditions at the brane positions. To this end, suppose the source
brane is not regarded to be arbitrarily thin and is instead
resolved to have a small proper width, $\ell$, within which some
new stress-energy turns on and smooths out the singular geometry
at $r = 0$.
\footnote{This kind of treatment has been recently used
in \cite{VC} to regularize other kinds of deformations of
the rugby ball geometry, in order
to accommodate on
the brane   an energy momentum tensor more general than
pure tension. See also \cite{branereg} for other proposals in this direction.
} Then it is the matching of this smooth
geometry internal to the brane with the external geometry which
decides which external solution is appropriate. This kind of
matching has been done explicitly for weakly-gravitating local
cosmic strings \cite{Ruth}, subject to suitable falloff conditions
for the interior energy density, with the result that these kinds
of objects choose the asymptotically conical solutions. We now
re-examine this matching within Einstein gravity for more general
strongly-coupled sources.

To do so suppose that the exterior metric of (\ref{bulkgeometry})
applies for $r \ge \ell$, and smoothly matches onto a regular
geometry for $0 \le r \le \ell$ of the same form
\be\label{intgeom}
    \exd s^2 =  {\cW}^{2}(r) \, \eta_{\mu\nu} \, \exd
    x^\mu \exd x^\nu +
    \exd r^2 + {\cA}^{2}(r) \, \exd\theta^2 \,,
\ee
since we again demand maximal symmetry for the infinite 4
dimensions. The explicit form of the coefficients $\cW$ and $\cA$
depends on the internal stress energy which resolves the structure
of the brane for $r \le \ell$. Requiring the interior metric to be
well defined at the origin implies
\be\label{limit0}
    \cW(0)=1 \,, \qquad \cW'(0)=0 \hskip1cm \hbox{and} \hskip1cm
    \cA(0)=0 \,, \qquad \cA'(0)=1 \,,
\ee
and the condition that it smoothly connect with the exterior
metric at $r = \ell$ implies:
\bea\label{limit0prime}
    &&\cW(\ell) = W(\ell) \approx \, W_0\,\ell^{\omega} \,,
    \quad \,\, \cW'(\ell) = W'(\ell) \approx \omega \, W_0\,
    \ell^{\omega-1} \nonumber \\
    &&\cA(\ell) = A(\ell) \approx A_0 \,\ell^{\alpha}
    \,, \qquad \cA'(\ell) = A'(\ell) \approx \alpha \,A_0 \,
    \ell^{\alpha-1}\,\,,
\eea
for some constants $W_0$, $A_0$, $\alpha$ and $\omega$.

There are two separate kinds of interpretational issues which this
kind of matching raises: Are the sources 3-branes or 4-branes?
And: If they are 3-branes how are their brane properties related
to the kind of external solution which is appropriate? We address
each of these in turn in the remainder of this section.

\medskip\noindent{\bf 3-Branes {\it vs} 4-Branes}

\medskip\noindent
In order for the source to be interpreted as a 3-brane, its proper
transverse size must be small compared with those of the exterior
space. Although this is generally true for the brane's proper {\it
radius}, since by definition $r > \ell$ for all points exterior to
the brane, in order to interpret the source as a 3-brane we also
demand it to be true for the brane's {\it circumference}, $C_b = 2
\pi A(\ell)$, relative to the circumference of circles, $C(r) = 2
\pi A(r)$, which surround the brane at some proper separation $r$.
(We could equivalently phrase this criterion in terms of the
transverse volume of the brane, $\Omega_\ell = \int_0^\ell dx \,
\cA(x)$, relative to the volume interior to various circles which
surround the brane, $\Omega(r) = \Omega_\ell + \int_\ell^r dx \,
A(x)$.)

Clearly, if $\alpha > 0$ then $A$ increases as $r$ increases and
so the circumference, $A(\ell)$, at the brane boundary is smaller
than the circumference of those circles which surround the brane
at radii $r > \ell$. This is what would be expected for a 3-brane
whose size is much smaller than the size of the bulk geometry
within which it sits. The opposite situation arises if $\alpha <
0$, since then $A(r)$ is a decreasing function as $r$ increases.
In this case the external geometry resembles that of the mouth of
a trumpet, with the circumference, $A(\ell)$, of the circle at the
boundary of the brane being in this case larger than the
circumference of all of the circles which surround the brane at $r
> \ell$. In this case the source brane is hard to interpret as a
point source within a larger bulk, and the bulk geometry,
(\ref{bulkgeometry}), is better interpreted as being sourced by a
4-brane than by a 3-brane.

For the single-brane solutions given by eq.~(\ref{bulkgeometry}),
the choices are $\alpha = 1$ or $\alpha = -3/5$, and so it is only
those geometries having conical singularities which would be
interpreted as being sourced by localized 3-branes according to
the above interpretation. The same is {\it not} true for the GGP
solutions, for which we have seen that $\alpha$ can be positive
but not equal to unity, and so furnish examples of geometries with
non-conical singularities for which $A(r)$ is an increasing
function of $r$. This possibility relies for its existence on the
presence of the dilaton in the bulk.

The very existence of these geometries raises the possibility that
there may be localized 3-brane sources for which the external
metric does not simply have the usual conical singularity. We now
try to ascertain precisely what properties of the brane, besides
its tension, dictate when and whether the external geometry can be
singular in this way.

\subsection{Smoothing Out 3-Branes}

To see what kinds of brane properties control the form of the
external metric, we specialize the above considerations to the
asymptotic forms obtained earlier for the singular GGP solutions
({\it i.e.} those having $\lambda_3$ nonzero).
As discussed in \S2 above, the change of variables
(\ref{chanofvar}) allows us in this case to explicitly compute the
constants $A_\pm$, $W_\pm$, $\omega_\pm$ and $\alpha_\pm$ defined by $W=W_\pm
r^{\omega_\pm}$ and $ A=A_\pm r^{\alpha_\pm}$. In the present case, near the
singularities
 at $\eta \to \pm \infty$ we have
\begin{eqnarray} \label{WAlimits}
 W_\pm^2 &=& \left(\frac{q\,\lambda_2}{4g\,\lambda_1} \right)^{\frac12}
  e^{\mp(\lambda_1\eta_1 -\lambda_2\eta_2)/2}
  \,\left[\frac{\Delta_\pm^2}{4} \left( \frac{\lambda_1\,
e^{\pm\lambda_1\eta_1}}{q}\right)^\frac12 \left(\frac{g\,
                    e^{\mp \lambda_2\eta_2}}{\lambda_2}  \right)^{5/2}
  \right]^{\omega_\pm} \\
  A_\pm^2 &=&  4\left(\frac{\lambda_1^3\lambda_2}{g\,q^3} \right)^{1/2}
                 e^{\pm(3\lambda_1\eta_1 + \lambda_2\eta_2)/2}
    \left[ \frac{\Delta_\pm^2}{4} \left( \frac{\lambda_1\,
  e^{\pm\lambda_1\eta_1}}{q}\right)^\frac12 \left(\frac{g\,
   e^{\mp \lambda_2\eta_2}}{\lambda_2}  \right)^{5/2} \right]^{\alpha_\pm}\,,
\end{eqnarray}
while the exponents $\omega_\pm$ and $\alpha_\pm$
are given by eq.~(\ref{omegaalpha}), which we reproduce here for
convenience:
\be \label{omegaalpha1}
    \omega_\pm = \frac{\lambda_2-\lambda_1}{5 \lambda_2
    -\lambda_1 \pm 2\lambda_3}  \qquad \hbox{and} \qquad
    \alpha_\pm
    = \frac{\lambda_2+3\lambda_1 \pm2\lambda_3}{5\lambda_2-\lambda_1
    \pm 2\lambda_3}\,.
\ee
and $\Delta_\pm=5\lambda_2-\lambda_1 \pm 2\lambda_3 $.
Recall also that $\alpha_\pm$ and $\omega_\pm$ satisfy the identity
$\alpha_\pm + 4 \omega_\pm = 1$. As
discussed earlier, in the limiting case $\lambda_1 = \lambda_2 =
\lambda$ (and so $\lambda_3=0$) these reduce to the case of a
conical singularity, for which
\be \label{limitcons}
    W_\pm=1\,, \hskip 1 cm  \omega_\pm=0\,, \hskip 1 cm
    A_\pm=\lambda\,, \hskip 1 cm \alpha_\pm=1\,\,,
\ee
but for $\lambda_3 \neq0$, the metric has a more complicated
curvature singularity at $r=0$.

Consider now the stress-energy which is responsible for smoothing
out the brane. We take for these purposes an internal
energy-momentum tensor having the most general
rotationally-invariant form
\be\label{intemt}
    {T^{M}}_{\,N} = - \hbox{diag}\left(\epsilon,\epsilon,\epsilon,\epsilon,
    p,\sigma\right) \,,
\ee
where the functions $\epsilon(r)$, $p(r)$ and $\sigma(r)$ vanish
for $r > \ell$. We ask  these to be related to the internal-metric
functions, $\cW(r)$ and $\cA(r)$, through the interior Einstein
equations, $G_{MN} = 8\pi G\, T_{MN}$, where we imagine here that
$T_{MN}$ also includes the energy-momentum of the bulk dilaton and
the gauge field within the brane.

The above expressions allow us to draw some preliminary
conclusions about the connection between the external geometry and
brane properties. For example, Einstein's equations allow us to
write the following expression for the brane tension
\begin{eqnarray}
   T_{\ell_\pm} &=& \int_0^{2\pi} \exd \theta \int_0^{\ell} \exd r
    \,\sqrt{-{g}_{6}} \; \epsilon= -\frac{1}{8\pi G }
                         \int_0^{2\pi} \exd \theta \int_0^{\ell}
    \exd r \, \cW^{4}(r) \, \cA(r) {G^t}_t
    \\
&=&  -\frac{1}{8\pi G }
                         \int_0^{2\pi} \exd \theta \int_0^{\ell}
\exd r\,\cW^{2}\,[3\,\cW'^2 \cA + 3\,\cW \,\cW'\cA'+3\,\cW\,\cW''\cA +\cW^2 \cA'']
                 \nn \\
&=& \frac{2 \pi}{8\pi G } \left[1- A_\pm\,W_\pm^4 \ell^{4 \omega +\alpha -1}
    (\alpha_\pm + 3 \omega_\pm) \right] 
+ \frac{2\pi}{8\pi G }\left(4\int_0^\ell \cW^{3} \cW' \cA' \,\exd r
    +6\,\int_0^{\ell} \cW^{2}
    \,(\cW')^{2} \,\cA \,\exd r
    \right) \nn \\
    &=& \frac{2 \pi}{8\pi G } \left[1- A_\pm\,W_\pm^4 (\alpha_\pm + 3 \omega_\pm) \right]
   + \frac{2\pi}{8\pi G } \left( 4\int_0^\ell
    \cW^{3} \cW' \cA' \,\exd r +6\,\int_0^{\ell} \cW^{2}
    \,(\cW')^{2} \,\cA \,\exd r
    \right)
     \,,\label{enthick}
\end{eqnarray}
where we have used the boundary conditions (\ref{limit0prime})
with $W_0=W_\pm$, $A_0=A_\pm$, $\alpha= \alpha_\pm$ and
$\omega=\omega_\pm$, corresponding to the constants associated
with the singularities at $\eta\to \pm\infty$ to evaluate the
integrals. The last line uses the identity $\alpha_\pm + 4
\omega_\pm = 1$. Notice that the integrals in the last line of
this expression tend to zero in the limit $\ell \to 0$, provided
that the metric functions $\cW$ and $\cA$ are nonsingular and
sufficiently well behaved throughout
the integration range $0 < r < \ell$. Consequently we find the
following general relation between the tensions, $T_\pm$, and the
parameters which govern the asymptotic form of the external
metric:
\be \label{Tformula}
    T_\pm = \lim_{\ell \to 0} T_{\ell_\pm} = \frac{1}{4 G }
    \left[1- A_\pm\, W_\pm^{4}\left(\alpha_\pm\, +
    3 \omega_\pm   \right)\right] \,.
\ee

If we use the explicit
expressions given earlier for $A_\pm$, $W_\pm$, $\alpha_\pm$ and $\omega_\pm$
as functions of the parameters $\lambda_1$, $\lambda_2$, $q$, {\it
etc.} into these expressions for the tensions, then we see that
eq.~(\ref{Tformula}) shows how to relate
the brane tensions to the parameters appearing in the bulk
solutions, even for the non-conical geometries. As is easily
verified, it reduces in the case of conical singularities to the
standard connection between the tension and defect angle, since in
this case inserting formulae (\ref{limitcons}) into
(\ref{enthick}) gives for the $\eta \to - \infty$ singularity
\be
    T = \frac{1}{4 G } (1-\lambda)\,.
\ee

We learn from this that the tension in itself is insufficient to
determine both of the powers $\alpha$ and $\omega$. So what is it
which determines whether the external geometry is unwarped and
conical or warped and more singular? A clue to this comes from the
$\theta-\theta$ components of the Einstein equations, which read
${G^\theta}_\theta = 8\pi G \, {T^\theta}_\theta$, or
\be
     10 \left( \frac{d}{dr} \ln \cW \right)^2
    + 4 \frac{d^2}{dr^2} \ln \cW = 8\pi G \,\sigma \,,
\ee
since this shows that a constant warp factor external to the
brane, $W'(\ell) = \cW'(\ell) = 0$, is only possible if $\sigma =
{T^\theta}_\theta$ also vanishes at $r = \ell$.
For instance $\sigma$ might be nonzero due to the microscopic
stress which resolves the brane, or it could be nonzero because of
the presence near the brane of nontrivial dilaton or
electromagnetic fields.
Clearly conical geometries must lie external to `thin wire' branes
for which ${T^\mu}_\nu = T \delta^2(x) \, {\delta^\mu}_\nu$ and
${T^r}_r = {T^\theta}_{\theta} = 0$, but it is the absence of
transverse stress-energy which is responsible for this fact.

In general, the small-$\ell$ limit is a singular one which depends
on the details of the microscopic physics which smooths out the
brane in question.\footnote{See, for instance, ref.~\cite{geroch}
for a discussion of many of the issues relevant to this limit.}
However, general physical arguments (see for instance
\cite{ELReviews,bdyELs}) ensure that for small $\ell$ it is always
possible to organize this dependence into an effective action
localized at the brane positions, with successive higher-dimension
terms suppressed by higher powers of $\ell$. For instance, we
speculate that it may be the appearance of effective operators
involving the normal components of the curvatures (such as $R_{ij}
= n_i^M n_i^M \, R_{MN}$, with $n_i^M, i=1,2$ denoting unit
normals to the 3-brane) within the localized brane action which
would reflect in this way the presence of non-zero $\sigma(\ell)$.
It would be of considerable practical interest to pin down this
issue, by finding the detailed matching between the properties of
the microscopic stress-energy profiles, $\epsilon$, $p$ and
$\sigma$, used above, and the corresponding effective operators in
the brane effective action.

\section{The Solutions of Ref.~\cite{Warped}.}

This section is dedicated to  demonstrate the equivalence between
the solutions of ref.~\cite{Warped} and the solution of
eq.~(\ref{ggp302}): that is, the general solution of
ref.~\cite{GGP} having two or fewer singularities that are {\it
conical}. We do so in order to clarify how the solutions of these
two references are related to one another.

\subsection{The Solutions and Their Relation to Those of
Ref.~\cite{GGP}}

This equivalence is proven by explicitly constructing the
coordinate transformation which relates the two solutions. The
solutions constructed in ref.~\cite{Warped} are given by~\footnote{We
present here a slight generalization of the solutions of ~\cite{Warped},
since we explicitly introduce an additive integration constant
$\phi_0$ for the scalar field.}
\begin{eqnarray}
    && \exd s_{6}^{2} = 2\,r[-\exd t^2 + \exd x_{3}^{2}]
            +h(r)\, \exd \psi^{2} +\frac{\exd r^{2}}{h(r)}
    \equiv   e^{{\phi}/{2}} \eta_{\mu \nu} \, \exd x^{\mu}\exd x^{\nu}
    \,+\, \exd s_{2}^{2} \,,\label{metr2} \\
    && \phi(r) =  2 \ln (2r) + \phi_0 \label{dil2}  \,,\\
    && F =\frac{\tilde{q}\,e^{-{\phi_0}/{4}}}{8\,r^3} \,
 \,\exd r \wedge \exd \psi  \,\label{gau2},
\label{gauf}
\end{eqnarray}
with
\begin{equation}\label{defh}
    h(r)= \frac{2m}{r} - \,g^{2}\,e^{-{\phi_0}/{2}}\,r
                      -\, \frac{\tilde{q}^{2}}{256\, r^3} \,.
\end{equation}
To obtain these expressions we take the solutions of
ref.~\cite{Warped} and make the replacements $\phi \to - \phi/2$
as well as re-scaling our units for the 6D Planck scale from
$\kappa^2 = 1$ to $\kappa^2= \frac12$, as is required in order to
conform with the conventions used here and in ref.~\cite{GGP}. We
also must take $R_{MN} \to - R_{MN}$ in order to change from the
Weinberg curvature convention used in \cite{Warped} to the MTW
conventions of \cite{GGP}. Finally, we also re-name the
integration constant ${\cal A}$ of ref.~\cite{Warped} to $\tilde
q$, according to $8 {\mathcal A}=\tilde{q}$. As written, this
solution has 3 integration constants, $\phi_0$, $\tilde{q}$ and
$m$, which is the same counting (before fixing the classical scale
invariance) as for the $\lambda_3 = 0$ solutions discussed above.

In order to show the equivalence of this solution with that of
eq.~(\ref{ggp302}), it is enough to change the  radial coordinate
$r$ in the solution (\ref{metr2}, \ref{dil2}, \ref{gauf}), to a new
coordinate $\eta$ via the relation
\be\label{defraseta}
    2 \,r \,e^{{\phi_0}/{2}}=\,W^2(\eta)
\ee
where the function $W(\eta)$ is the one given in formula
(\ref{genWformula}). From here, it is clear that the scalar $\phi$
of eq.~(\ref{dil2}) matches the scalar of eq.~(\ref{ggp300}) in
the solution of \cite{GGP}. Using this definition of $\eta$ leads
to the following formula
\be
    \exd\,r= e^{-\phi_0/2} \frac{\lambda W^2}{4} \Big\{ \tanh{[\lambda \left(
    \eta-\eta_1\right)]} - \tanh{[\lambda \left( \eta-\eta_2\right)]}
    \Big\}\,\exd\,\eta\,.
\ee

At this point, we match the integration constants of the solution
in \cite{Warped} with the integration constants of the solution in
\cite{GGP}. In particular, we must re-write the integration
constants $q$ and $\lambda$ of the solution in section
(\ref{sec:properties}) in terms of the integration constants given
here by
\be
    \tilde{q}^{2}= q^2 \, e^{-{5 \phi_0}/{2}}\,,
\ee
and we set
\bea
    \frac{1}{\lambda} \sinh{[\lambda(\eta_2-\eta_1)]} &=& \frac{2\,
    e^{{\phi_0}/{2}}}{gq}\,
    \,, \nn\\
    \label{condrel}
    \\
    \cosh{[\lambda(\eta_2-\eta_1)]} &=& \frac{16m\,e^{{3 \phi_0}/{2}}
    }{g q}
    \,.\nn
\eea
Substituting the definition of $r$ in terms of $\eta$, given in
(\ref{defraseta}), into the solutions (\ref{metr2}, \ref{dil2},
\ref{gauf}, \ref{defh}), and using the conditions (\ref{condrel}),
it is straightforward to show that one obtains in this way the
solution discussed in \S\ref{sec:properties}.

Notice that the change of variables defined by
eq.~(\ref{defraseta}) is only locally well-defined. This is
because it defines $r$ to be a double-valued function of the
coordinate $\eta$, since $W$ increases as one moves away from the
brane at $\eta \to -\infty$ and then decreases again as the other
brane is approached as $\eta \to +\infty$. In particular, the
special case of the unwarped rugby-ball solutions can only be
obtained from the warped solutions of this section through an
appropriate limiting procedure -- as we show in detail below. This
limiting procedure is required because in the special case of the
unwarped rugby-ball solutions the change of variables
(\ref{defraseta}) is singular, since for these solutions $W$ is a
constant.

This establishes the equivalence of the solutions of
ref.~\cite{Warped} with those solutions of ref.~\cite{GGP} for
which $\lambda_3 = 0$. The integration constants, $q$, $\lambda$
and $\eta_1-\eta_2$ of one solution are interchangeable for the
integration constants, $m$, $\tilde{q}$, and $\phi_0$ of the
other. In particular, the choice $\lambda=1$ which produces a
solution having only a single conical singularity corresponds to
adjusting the choice of the parameter $m$ in terms of $\tilde{q}$
and $\phi_0$ in such a way as to make one of the tensions vanish.

\subsection{Topological Constraints}\label{tc2}

Having established the connection between the solutions of the two
papers, it is possible to use directly the results of
ref.~\cite{Warped} to infer which brane properties are consistent
with the constraints, such as those from topology. Physically, we
are in principle free to choose the tensions, $T_\pm$, of each of
the source branes\footnote{We assume here the absence of the
magnetic coupling between the 3-branes and the background magnetic
field. As discussed in refs.~\cite{Towards} and \cite{Warped} the
inclusion of these couplings both introduces two new physical
parameters --- the magnetic charges of each brane
--- and modifies the topological constraint \cite{Warped} on the
magnetic flux.} as well as choose the total monopole number, $N$,
of the background magnetic flux. The topological constraint,
eq.~(\ref{quaco}), in terms of the coordinates in ref.~\cite{Warped},
reads
\be
 \label{topcond2}
   \frac{ \tilde q}{32} \left( \frac{1}{r_-^2} - \frac{1}{r_+^2} \right)
     = \frac{N}{\tilde{g}\, e^{-\phi_0/4}} \,,
\ee
where $N$ is an integer, and as before $\tilde{g}$ is the gauge
coupling appropriate to the background gauge field. In this
expression the quantities $r_\pm$ denote the combinations
\be \label{rpmdefs}
    r^2_\pm = \frac{m}{g^2\, e^{-\phi_0/2}}\left[ 1\pm \sqrt{1-
                 \left(\frac{g\, \tilde q\, e^{-\phi_0/4}}{16\,m}
                 \right)^2}\right]
                 \,,
\ee
which are defined by the roots of the function $h(r)$ which
appears in eq. (\ref{defh}), and as such correspond to the
positions of the conical singularities. Ref.~\cite{Warped} shows
that they are also related to the tensions of the branes at the
singularities by
\be \label{expten2}
    T_{\pm}=4\pi \left[1-\frac{g^{2}e^{-\phi_0/2}}{2
    r_{\pm}^{2}}(r_+^{2}-r_-^{2})\right]\,.
\ee
In writing this result we must keep in mind the change of
convention from $\kappa^2 = 1$ to $\kappa^2 = \frac12$ between
ref.~\cite{Warped} and here, and the convention introduced there
that the tensions of the two branes are labelled such that $T_+
\ge T_-$. More general units are obtained by the replacement
$T/4\pi \to \kappa^2 T/2 \pi = 4G \, T$.

Using eqs.~(\ref{rpmdefs}) and (\ref{expten2}) in (\ref{topcond2})
allows the topological constraint to be written directly in terms
of the tensions and gauge couplings as follows:
\be \label{tensiontopology}
    \frac{g^2 e^{-\phi_0/2}}{2} \left( \frac{T_+ - T_-}{4 \pi}
    \right) = N^2 \left( \frac{g^2}{\tilde{g}^2} \right) \,.
\ee

The above expressions degenerate in the limit of equal tensions,
since on one hand the positivity of $h(r)$ requires $r_- < r <
r_+$ and on the other hand the equal-tension condition, $T_+ =
T_-$, implies $r_+ = r_-$. This degeneracy is consistent with the
singular nature of the change of variables (\ref{defraseta}) in
this case. Notice, however, that if $g^2 e^{-\phi_0/2} \to \infty$
as $r_- \to r_+$, in such a way as to ensure that the product $g^2
e^{-\phi_0/2}(r_+^2 - r_-^2)$ approaches a finite constant $2 k\,
r_+^2$, then eq.~(\ref{expten2}) implies $T_\pm \to 4\pi(1 - k)$.
Furthermore, using this limit in the identity $T_+ - T_- = 2\pi
g^2 e^{-\phi_0/2} (r_+^2 - r_-^2)^2/(r_+^2 r_-^2)$ shows that $g^2
e^{-\phi_0/2}(T_+ - T_-) \to 8 \pi \, k^2$, and so in this case
the topological constraint, (\ref{tensiontopology}), reduces to
$k^2 = N^2 (g^2/\tilde{g}^2)$.

We see in this way how the unwarped solution may be retrieved from
the warped solution through an appropriate limiting
procedure.\footnote{We thank Jim Cline for conversations on this
point.}

\subsection{Tension Constraints}

An important property of these solutions is that the tensions of
the two branes may not be chosen independently, since they are
implicitly related to one another by means of the formulae
presented above. This implicit relation between the tensions was
first found in Ref.~\cite{Warped}, and can be made explicit by
eliminating $r_+/r_-$ from eq.~(\ref{expten2}), giving
\be \label{Tconstraint}
    \frac{g^2\, e^{-\phi_0/2}}{2} \left( \frac{T_+ - T_-}{4 \pi}
    \right) =  \left(1 -
    \frac{T_+}{4\pi} \right) \left( 1 - \frac{T_-}{4\pi} \right)
    \,.
\ee
This generalizes the equal-tension constraint which is required
for the existence of the unwarped rugby-ball solution to those
warped solutions having only conical singularities. Since the
solutions of ref.~\cite{Warped} are equivalent to those of
ref.~\cite{GGP} having conical singularities, we see that conical
singularities are only possible for a one-parameter subset of the
$T_+ - T_-$ plane.

Two consequences of eq.~(\ref{Tconstraint}) provide useful checks.
First, notice that simplifying the left-hand side of
eq.~(\ref{Tconstraint}) using eq.~(\ref{tensiontopology}) leads
directly to eq.~(\ref{GGPTensionconstraint}), which expresses the
tension constraint for the $\lambda_3 = 0$ solutions of
ref.~\cite{GGP}. Second, since $g^2 e^{-\phi_0/2}(T_+ - T_-) \to 8
\pi \, k^2$ in the equal-tension limit, eq.~(\ref{Tconstraint})
reduces in this limit to $(1 - T/4 \pi)^2 \to k^2$, as it must.

Eqs.~(\ref{Tconstraint}) and (\ref{tensiontopology}) also make it
particularly easy to see what must be done in order to ensure that
one of the conical singularities vanishes, since this is
accomplished by setting one of the tensions to zero. It is in
particular clear that there are two ways to do so: ($i$) set $T_+
= 0$ and leave a single brane having negative tension, $T_-$; or
($ii$) set $T_- = 0$ and leave a single brane having positive
tension, $T_+$. These correspond to the choice, $\lambda = 1$,
considered earlier when discussing the GGP solution.

Now, we saw in previous sections that having positive tension was
only possible when the background-field gauge coupling,
$\tilde{g}$, is larger than the coupling, $g$, which appears
directly in the dilaton potential. We now use
eqs.~(\ref{Tconstraint}) and (\ref{tensiontopology}) to see how
these same constraints emerge. To this end, using $T_- = 0$ in
eq.~(\ref{Tconstraint}) immediately leads to the conclusion
\be\label{tpos}
    \frac{T_+}{4\pi} = \frac{2}{2 + g^2 e^{-\phi_0/2}}\,,
\ee
which confirms that the tension of the remaining brane is
positive. Using this in eq.~(\ref{tensiontopology}) then leads to
the following expression:
\be\label{sectopcond}
    \frac{\tilde{g}^{2}}{g^{2}}\,\frac{1}{N^{2}} =
    1+\frac{2e^{\phi_0/2}}{g^{2}} \,.
\ee

This is the key formula. Consider first the case where the
background gauge field is in precisely the $U(1)_R$ gauge
direction, and so $\tilde{g}^{2} = g^{2}$. In this special case
the left-hand side of the above relation is always less than one,
while the right-hand side is always bigger than one, showing that
the choice $g = \tilde{g}$ makes it impossible to choose $T_- = 0$
(in agreement with the discussion of Sections (\ref{tc1})).
However, as we also saw in Section (\ref{tc1}), we can instead
choose $\tilde{g} > g$, and in this case it is possible to satisfy
the quantization condition (\ref{sectopcond}). This confirms that
a brane with positive tension as in formula (\ref{tpos}) is indeed
compatible with all of the constraints (provided that both of the
couplings, $\tilde{g}$ and $g$, are not too small).

\section{Discussion}

We close with a summary of our results, and a discussion of what
they can mean for the 6D self-tuning issue.

\subsection{Summary}

In this paper we explore some aspects of the general solutions to
6D chiral supergravity which were first given in ref.~\cite{GGP},
focusing on the singularity structure of these solutions.

In particular we show that only a subset of these solutions --
those for $\lambda_3 = 0$ -- have purely conical singularities.
These solutions, moreover, are exactly equivalent to the warped
brane solutions given in ref.~\cite{Warped}, as we show by
explicitly constructing the coordinate transformation which
relates them. We show explicitly that these geometries can be
consistent with both positive and negative tension branes,
depending on how the background gauge field is oriented inside the
complete gauge group. The ability to find configurations with only
positive tension branes, resolves an apparent discrepancy between
\cite{GGP} and \cite{Warped}. This observation is interesting
since the existence of geometries with a pair of positive tension
branes is one of the motivations for considering codimension two
brane-world in six dimensions. Moreover it can also be important
for further studies of the stability properties of these
configurations.

For those solutions not having conical singularities --- \ie\ for
$\lambda_3 \neq 0$ --- the metric near these singularities behaves
as $\exd s^2_2 \sim \exd r^2 + A^2 r^{2\alpha} \, \exd \theta^2$
as $r \to 0$, for appropriate constants $A$ and $\alpha$. The warp
factor also vanishes in this limit as $W \sim r^\omega$, with
$\omega > 0$. We find that there are solutions for which $\alpha$
is positive for both singularities, and other solutions for which
$\alpha$ is positive at one singularity but is negative at the
other. There are no cases where $\alpha < 0$ for both
singularities.

The cases with $\alpha < 0$ resemble the behavior of known
single-brane solutions to pure Einstein gravity (without the
dilaton), which we argue are more likely to be interpretable in
terms of the fields produced by 4-brane sources than by 3-brane
sources. We base this interpretation on the fact that the
circumference of the circles enclosing the brane shrink as one
moves away from it, making it difficult to think of the brane as a
much smaller object than the surrounding bulk space. We regard the
further exploration of the properties of these solutions to be
very interesting, inasmuch as they open up a new class of bulk
geometries which may have interesting phenomenological properties.

Situations where $\alpha > 0$ more resemble the geometry outside
of a small codimension-2 object, and so naturally lend themselves
to an interpretation in terms of 3-brane sources. The geometries
in this case have a curvature singularity at the origin which is
only conical in the special situation $\alpha = 1$. We derive an
expression for the general solution which relates the tension of
such a 3-brane source to the asymptotic behavior of the bulk
fields near the brane. When specialized to conical singularities
this expression reduces to the usual formula relating the brane
tension to the size of the conical defect angle.

What is puzzling in this instance is identifying which properties
are required of the source branes to produce the new kinds of bulk
geometries instead of the usual conical geometries which are
familiar from weakly-gravitating systems. Our preliminary analysis
indicates that this could be due to the microscopic physics which
resolves the brane structure, or due to the nontrivial behavior of
other bulk fields (like the dilaton) near the brane, provided
these lead to nonzero normal components to the underlying
microscopic stress-energy tensor. A more explicit characterization
of the matching conditions which connect between the brane
properties which are responsible and the asymptotic forms taken by
the bulk metric would be of considerable interest, which we leave
for future work.

\subsection{Potential Significance for Self Tuning}

The properties of these exact solutions also have a potential
relevance for the 6D self-tuning proposal of ref.~\cite{Towards}.
In particular, as was discussed in ref.~\cite{Update}, if this
proposal is to be successful it must ultimately explain why the
small effective 4D cosmological constant seen by brane observers
is robust against arbitrary changes to the various brane tensions,
such as might occur due to phase transitions on the branes.

Imagine, then, we start off with one of the geometries having
conical singularities (such as the rugby ball, for example), for
which brane observers experience a flat 4 dimensions. Due to the
considerations of \S4, we know the solutions of ref.~\cite{Warped}
describe the most general such geometry, and it follows that the
two brane tensions must satisfy the constraint relation,
eq.~(\ref{Tconstraint}). Suppose also that at time $t=0$, one
of these tensions, say $T_-$, is instantaneously locally perturbed
to a new value, $T'_-$, and then held fixed. We imagine that the
other tension, $T_+$, does not change during this process.

Since in general the new tensions, $T_+$ and $T'_-$, do not
satisfy eq.~(\ref{Tconstraint}), the bulk geometry after this
tension change can no longer be described by the solutions of
refs.~\cite{GGP,Warped}. Locality and conservation laws (energy
and angular momentum) dictate that the bulk fields will experience
a transient time-dependence as the system tries to find an
equilibrium configuration consistent with its new tensions. Short
of knowing the explicit time-dependent solutions, what can be said
about the late-time equilibrium configuration to which the system
might go?

There are a few conclusions which can be drawn, even given only
knowledge of the solutions of ref.~\cite{GGP}.

\smallskip\noindent$\bullet$
First, since time evolution is continuous, the two topological
constraints (monopole number and Euler number \cite{Towards,GGP})
of the initial configuration are guaranteed to be preserved during
the evolution. Of course, the particular expressions of these
constraints as functions of the parameters of the initial unwarped
solutions in general must change, since the new solutions cannot
be unwarped for arbitrary final tensions. But the topological
nature of the constraints ensures that they automatically remain
satisfied by the final solutions, as well as the time-dependent
solutions which lead to them. Similarly, conserved quantities like
energy and angular momentum must also remain unchanged during the
time-dependent evolution.

\smallskip\noindent$\bullet$
Second, {\it if} it were known that for any choice of conserved
(including topological) quantities and for any pair of tensions
there existed a solution within the class of solutions of
ref.~\cite{GGP}, then we might reasonably expect that the endpoint
of the time-dependent evolution would be this static solution,
corresponding to the various KK modes all settling down into new
minima of their potentials. Once there we would know the brane
observers would again experience flat 4D space because all of
these solutions have this property. This makes it of central
importance to understand how the bulk geometries of the general
GGP solutions are related to brane tensions which source them.

Given our present knowledge, we {\it can} say something about
whether or not solutions exist for pairs of tensions $T'_-$ and
$T_+$, at least for those which are small perturbations of a
geometry having conical singularities. There are two cases: either
these new tensions satisfy the constraint of
eq.~(\ref{Tconstraint}) or they do not. If they do, then there
exists a new solution which again only has conical singularities,
and this is most likely the endpoint towards which the time
evolution leads.

If the final tensions do not satisfy eq.~(\ref{Tconstraint}), then
any static solution to which the system tends cannot have just
conical singularities. We then need to know whether there exist
solutions with $\lambda_3 \ne 0$ for the given tensions which are
consistent with the initial conserved quantities. If so, then
again we could expect that these solutions would be the endpoints
of the system's evolution, and so that the brane observers would
eventually experience flat 4D space after any transient
time-dependence passes.

Happily, we can begin to address this issue using formulae
(\ref{WAlimits}), (\ref{omegaalpha1}) and (\ref{Tformula}), which
relate the brane tensions, $T_+$ and $T_-$, to the solution
parameters, $\lambda_1$ and $\lambda_2$, $\eta_1$, $\eta_2$, $q$.
These formulae are useful, since they (in principle) allow us to
infer whether or not there exists a choice for the parameters
$\lambda_1$, $\lambda_2$ {\it etc.} which correspond to an
arbitrary choice of brane tensions. To decide this we must see
whether these formulae are invertible, because if they are then
they can (at least locally) be solved for $\lambda_1$ and
$\lambda_2$ as functions of $T_\pm$. In order to see how this
works in detail, we consider the simplest case $\eta_1=\eta_2=0$
and $4g=q$ in (\ref{WAlimits}), (\ref{omegaalpha1}) and
(\ref{Tformula}).

The invertibility of the relations relating the tensions to the
bulk-solution parameters are locally ensured if the determinant
\be
    J = \det \left(%
    \begin{array}{cc}
    \partial T_+/\partial \lambda_1 & \partial T_+
            /\partial \lambda_2  \\
    \partial T_-/\partial \lambda_1 & \partial T_-
            /\partial \lambda_2 \\
    \end{array}%
    \right) \ne 0 \,.
\ee
However, it is easy to see that
the tensions can be written as:
$T_+ = 1 - \lambda_2 \, F_+(u)$ and $T_- = 1 -
\lambda_2 \, F_-(u)$, where $F_+$ and $F_-$ are known functions of
$u= \lambda_1/\lambda_2$ and we suppress the dependence on the
parameter $q$, we find
\be
    J \propto  \frac{d}{du} \left[ \frac{F_+}{F_-} \right] \,.
\ee
Now, there must exist a choice for $\lambda_1 =
\lambda_1(T_+,T_-)$ and $\lambda_2 = \lambda_2(T_+, T_-)$ in the
immediate vicinity of any $T_\pm$ for which $J \ne 0$, and so this
makes the existence of a solution for any such $T_\pm$ quite
plausible. (Notice also that $J = 0$ for the conical solutions --
as it must -- since for these the quantities $F_+$ and $F_-$
become independent of $\lambda_1$ and $\lambda_2$.).

Of course, it still remains to impose the topological constraints,
which we imagine using to determine the remaining parameters,
$\eta_1$, $\eta_2$ and $q$ of the solution. (At first sight it
appears that we are not free to choose $q$ in this way, due to the
condition $q = 4g$ which follows from our simplifying assumption
that $\eta_1=\eta_2=0$. Recall, however, that because $\lambda_3
\ne 0$ we may rescale $g$ at will, through a rescaling under which
$\eta_i$ shifts. In this way we see that it is natural that
a solution satisfying all of the constraints exists for any pair
of given tensions.)

We are encouraged by this to expect that if the initial tensions
were to be perturbed, then the time dependence initiated by this
perturbation would ultimately lead to a new solution within the
GGP class, for which the intrinsic curvature again vanishes. This
is precisely what self-tuning is meant to do.\footnote{Notice that
if the resulting solution should be heavily warped, then this can
affect the success of the quantum part of the self-tuning argument
\cite{Update}, but a quantitative discussion of this issue goes
beyond the scope of the present paper.} We leave a careful
analysis of this possibility for future work.

\section*{Acknowledgements}

We thank J. Garriga and M. Porrati for an interesting e-mail
exchange, and Y. Aghababaie, Z. Chacko, J. Cline, G. Gibbons, G. Moore, S.
Parameswaran and J. Vinet for stimulating discussions about
self-tuning in 6D supergravity. C.B.'s research is supported by
grants from NSERC (Canada), FQRNT (Qu{\'e}bec) and McGill
University. F.Q. is partially supported by PPARC and a Royal
Society Wolfson award. G.T. is supported by the European TMR
Networks HPRN-CT-2000-00131, HPRN-CT-2000-00148 and
HPRN-CT-2000-00152. I.Z. is  supported  by the United States
Department of Energy under grant DE-FG02-91-ER-40672.

\section{Appendix: The Garriga-Porrati Analysis}

In this appendix we briefly discuss a recent paper by Garriga and
Porrati, \cite{GP}, which re-examines self-tuning in 6 dimensions,
both for theories with and without supersymmetry. The main claim
of this paper is that simple arguments can be used to show that
the {\it classical} part of the self-tuning can already be seen to
fail in both cases. For the non-supersymmetric case this claim
supports earlier, more detailed, studies
\cite{ignacio,VC,nilles,GKW} who also find difficulties with
self-tuning within the non-supersymmetric context. It also agrees
with the analysis of ref.~\cite{Warped}, which identifies a
classical scale invariance as playing an important role in
ensuring the intrinsic 4D geometries to be flat. The 6D
supersymmetric field equations enjoy such a scale invariance,
which is not present in the non-supersymmetric case because of the
necessity there to introduce a bulk cosmological constant in order
to stabilize the extra dimensions.

Our purpose in this appendix is to show why the arguments of
ref.~\cite{GP} do not suffice to establish their conclusion for
the supersymmetric theories. In particular, their arguments rely
on the assertion that reliable information about the solutions of
the full 6D equations may be obtained by studying the solutions to
a system of 4D equations obtained by truncating the 6D equations
about a particular ansatz. Although this kind of reasoning can
sometimes work in Kaluza Klein reductions, it need not and does
not in the compactification of 6D supergravity which is of
interest here and in ref.~\cite{GP}. We emphasize that the
ultimate success of the 6D self-tuning mechanism remains open, but
that a decisive test requires a more careful look at the solutions
of the full 6-dimensional field equations (which is in progress).

The key assumption of their analysis is the ansatz given by their
eqs. (3) and (6), which includes a dilaton, $\phi = \phi(x)$, and
\be  \label{GP1}
    \exd s^2 = e^{-2\psi(x)} \, g_{\mu\nu}(x) \, \exd x^\mu \exd
    x^\nu + e^{2 \psi(x)} \, \exd \Sigma^2_\lambda \,,
\ee
where
\be \label{GP2}
    \exd \Sigma^2_\lambda = \exd \theta^2 + \lambda^2 \sin^2\theta
    \, \exd \varphi^2 \,,
\ee
is the metric on the rugby-ball geometry constructed by removing a
wedge of defect-angle $\lambda$ from the 2-sphere as in equation
(\ref{defect}). The defect angle is proportional to the tension of
each of the two branes which source this solution, which must
necessarily be equal to one another. The background Maxwell field
also is taken proportional to the 2D volume form, $\omega$,
according to $F = b(x) \, \omega$. This ansatz could be motivated
by the knowledge that it includes the solutions of
ref.~\cite{Towards} as the special case $g_{\mu\nu} =
\eta_{\mu\nu}$ with $b$, $\phi$ and $\psi$ all constants. These
authors impose the requirements of flux conservation, $\exd F =
0$, which carries within it the local information that underlies
the global statement of the flux quantization described for this
solution in ref.~\cite{Towards}.

The argument proceeds by using this ansatz to truncate the 6D
action down to 4D, leading to the 4D action (their equation (18))
\be
    S = c \int d^4x \; \sqrt{-g} \Bigl[ R(g) - (\partial \phi)^2 -
    4 (\partial \psi)^2 - V_\alpha(\phi,\psi) \Bigr] \,,
\ee
for some constant $c$ and scalar potential $V$. The properties of
the solutions to the 4D equations following from this action are
then discussed, from which they draw their conclusions.

One of these conclusions involves asking precisely the right
question (also asked in ref.~\cite{Update}): What would happen to
the system if one were to take one of the tensions in the
rugby-ball solution and change it, so the two source branes no
longer have equal tensions? In the non-supersymmetric case the
authors state what they think the new static solution would be
like, after any transient time-dependence has passed: they expect
two rugby-ball like domains surrounding each of the branes, which
match into one another through a thin domain wall. For the
supersymmetric case they predict a runaway wherein the fields
$\psi$ and $\phi$ roll out to asymptotic values. In either case it
is claimed that this argument suffices to rule out the possibility
of successful self-tuning, already at the classical
level.\footnote{These arguments are similar to those which led
people to believe in the 1980's that the solutions to 6D
supergravity having monopole number not equal to $\pm 1$ must be
curved in 4D. The surprise with the solutions of ref.~\cite{GGP}
was that this curvature arises as warping and not due to curvature
of the {\it intrinsic} 4D geometry.}

Implicit in these conclusions is that the original ansatz for the
metric, eqs.~(\ref{GP1}) and (\ref{GP2}), remains unchanged with
only $\phi$ and $\psi$ varying in 4D space and time. This amounts
to assuming that the excitation of modes of the full 6D theory in
directions away from their ansatz are not excited. One test of
this assumption can be made by comparing to the exact 6D solutions
of refs.~\cite{GGP,Warped}, discussed here,  since these include
explicit 6D field configurations which are appropriate to branes
having different tensions (although as pointed out here, and in
\cite{Warped}, these tensions cannot be arbitrary). In particular,
these solutions are static and also have a flat 4D geometry, just
as for the original rugby ball, a circumstance which is missed by
the 4D arguments of ref.~\cite{GP}.

The reason these exact solutions are missed is that they do not
satisfy the ans\"atze ~(\ref{GP1}) and (\ref{GP2}). In particular
the 4D metric and dilaton are nontrivially warped over the extra
dimensions ({\it i.e.} both $\psi$ and $\phi$ acquire nontrivial
dependence on the coordinate $\theta$). This is not a big surprise
because the development of a dependence on the 2D coordinates like
$\theta$ corresponds to the excitation of some of the KK modes in
these directions. The crucial point is that this is {\it always}
possible because the typical mass of these KK modes, $m_{KK}$, is
precisely the same as the mass of that combination of $\psi$ and
$\phi$ which is not the scale invariance dilaton
\cite{abpq1,GibbonsPope}. As soon as there is sufficient energy to
excite nontrivially both $\phi$ and $\psi$ independently, there is
necessarily also enough energy to excite nontrivially generic KK
modes. This is why the KK modes cannot be integrated out to allow
a simple 4D analysis, and so shows why the system's response to
changes of tension is intrinsically a 6D problem. It is only the
evolution of the KK zero mode(s) which generically can be followed
by a simple 4D truncation.\footnote{Ref.~\cite{GP} also appeal to
Weinberg's no-go theorem, but see ref.~\cite{Update} for a
discussion of why the 6D proposal can evade this no-go result.}

As we emphasize here and elsewhere \cite{Update}, studying how the
full 6D dynamics responds classically to changes of tension is an
extremely interesting question, which is central to establishing
the validity or not of the 6D self-tuning proposal. Establishing
the size of quantum corrections is equally important
\cite{Update}. Fortunately or unfortunately, the present reports
of its demise --- like those of Mark Twain in an earlier time ---
remain premature.

\end{document}